\documentclass[twocolumn,pre,showpacs,floatfix]{revtex4}

\usepackage{bm}
\usepackage{amsmath}
\usepackage{amsbsy}
\usepackage{amsthm}
\usepackage{amssymb}
\usepackage{latexsym}
\usepackage[dvips]{color,graphicx}

\newcommand{\be}{\begin{equation}}
\newcommand{\ee}{\end{equation}}
\newcommand{\bea}{\begin{eqnarray}}
\newcommand{\eea}{\end{eqnarray}}

\begin{document}

\title{Numerical Linked-Cluster Algorithms.
II. $t$-$J$ models on the square lattice}

\author{Marcos Rigol}
\affiliation{Department of Physics and Astronomy, 
University of Southern California, Los Angeles, California 90089, USA}
\author{Tyler Bryant}
\affiliation{Department of Physics, University of California, Davis,
California 95616, USA}
\author{Rajiv R.~P.~Singh}
\affiliation{Department of Physics, University of California, Davis,
California 95616, USA}

\date{\today}

\pacs{05.50.+q,05.70.-a,75.10.Lp,05.10.-a}

\begin{abstract}
We discuss the application of a recently introduced numerical 
linked-cluster (NLC) algorithm to strongly correlated itinerant models. 
In particular, we present a study of thermodynamic observables:
chemical potential, entropy, specific heat, and uniform susceptibility for 
the $t$-$J$ model on the square lattice, with $J/t=0.5$ and $0.3$. 
Our NLC results are compared with those obtained from high-temperature 
expansions (HTE) and the finite-temperature Lanczos method (FTLM). 
We show that there is a sizeable window in temperature where NLC 
results converge without extrapolations whereas HTE diverges. 
Upon extrapolations, the overall agreement between NLC, HTE, and FTLM 
is excellent in some cases down to $0.25t$. At intermediate temperatures
NLC results are better controlled than other methods, making it easier to 
judge the convergence and numerical accuracy of the method.
\end{abstract}

\maketitle

\section{Introduction \label{introduction}}

In a recent paper \cite{rigol06} we introduced a   
linked-cluster algorithm, which we called the numerical 
linked-cluster (NLC) algorithm, that allows one to obtain 
temperature-dependent properties of quantum lattice models 
in the thermodynamic limit from the exact diagonalization of 
small clusters. A detailed exposition of NLC and its application 
to quantum spin models on square, triangular, and kagom\'e 
lattices has been presented in Ref.\ \cite{rigol06a}. There we 
have shown that for many spin models and thermodynamic quantities 
NLC results can be substantially more accurate than HTE and 
exact diagonalization (ED).

In this paper we discuss how to use NLC to calculate properties of 
strongly correlated itinerant models. In particular, we study the 
thermodynamics of the planar $t$-$J$ model on the square lattice.
This model was introduced by Anderson and others \cite{anderson87,zhang-rice} 
as a means to understanding the microscopic mechanism for 
high-temperature superconductivity \cite{bednorz86,wu87}. 
It is one of the simplest models that allows one 
to study the interplay between itinerancy of electrons and their
spin fluctuations, possibly leading to superconductivity and
many other exotic quantum phases \cite{chakravarty}.

In spite of its simplicity, understanding finite-temperature 
thermodynamic properties of the $t$-$J$ model has proven to be
a very challenging task \cite{jaklic00}. Quantum Monte Carlo 
simulations suffer from severe sign problems, which become a
major difficulty at low temperatures. The two general approaches 
that have been commonly used to study this model are ED and HTE. 
ED studies in which one fully diagonalizes the $t$-$J$ Hamiltonian 
are difficult since they can only be done for very small systems 
\cite{sokol93,tsunetsugu97}, as a consequence of which finite 
size effects are very large. A more efficient approach to 
this problem is the finite-temperature Lanczos method (FTLM), 
which has been developed by Jakli\v c and Prelov\v sek (JP) \cite{jaklic94}.
Within this approach the full thermodynamic trace is reduced 
by randomly sampling the eigenstates of the Hamiltonian. 
This allows one to study larger systems sizes in an unbiased way,
but still finite size effects become relevant as the temperature
is lowered.

In order to obtain results in the thermodynamic limit one can use 
high-temperature expansions (HTE) 
\cite{putikka92,singh92,shastry93,putikka94}. 
Within this method the properties 
of the system are expanded in powers of the inverse temperature 
$\beta$ \cite{domb,book}. These expansions, carried out to order 
$\beta^N$, where $N$ is of order 10, 
provide accurate numerical results within the radius of convergence 
of the series. The temperature scale where HTE converges
can be rather high for $t$-$J$ models. It is typically set by $t$
when $t>J$. Beyond the region of convergence, series extrapolation 
methods \cite{guttmann} allow one to calculate thermodynamic properties, 
but their reliability remains uncertain. Because the region of convergence
is small in inverse temperature, extrapolations even to temperatures 
of order $J$ are more sensitive to the choice of the extrapolation method 
and variables and hence less reliable than for purely spin models.

Here, we study thermodynamic properties of the $t$-$J$ model using NLC. 
The basic idea is to show with examples the advantages and disadvantages
of NLC as compared with FTLM and HTE, and the region of temperatures 
that within NLC can be accessed for the different observables of interest. 
The exposition is organized as follows. In Sec.\ \ref{density} we introduce 
the $t$-$J$ Hamiltonian and discuss the basic ideas of the NLC calculation
for itinerant models. The calculations must be done in the grand
canonical ensemble and the change from chemical potential to density
must be done numerically. We will show that this works quite well.
In the remaining sections we present results for 
the entropy (Sec.\ \ref{entropy}), the uniform susceptibility 
(Sec.\ \ref{suscept}), and the specific heat (Sec.\ \ref{cv}). 
The conclusions are presented in Sec.\ \ref{conclusions}

\section{Grand Canonical NLC \label{density}}

The $t$-$J$ Hamiltonian can be written as
\begin{eqnarray}
{\cal H}&=&-t\sum_{\langle i,j\rangle,s} 
P\left(c^\dagger_{is}c^{}_{j\sigma} + \mathrm{H.c.} \right)P \nonumber \\
&&+J \sum_{\langle i,j\rangle} 
\left({\bf S}_i\cdot{\bf S}_j-\frac{1}{4}n_i n_j \right),
\label{tJ}
\end{eqnarray}
where $c^\dagger_{is}$ and $c^{}_{is}$ are the creation
and annihilation operators for an electron with spin 
$s= \uparrow, \downarrow$ on a site $i$, 
$n_i=\sum_{s}c^\dagger_{is}c^{}_{is}$ is the 
density operator, $P$ is a projection operator to ensure 
no hopping produces doubly occupied sites, i.e., we assume 
that the local Coulomb repulsion is very large such that 
two electrons (with antiparallel spin) cannot be on the same 
lattice site, and 
\begin{equation}
{\bf S}_i=\frac{1}{2}\sum_{ss'} 
c^\dagger_{is}{\bf\sigma}_{ss'} c^{}_{is'}
\end{equation}
is the local spin operator (${\bf\sigma}$ are the Pauli 
matrices). The sums $\langle i,j\rangle$ in Eq.\ (\ref{tJ})
run over nearest-neighbor sites.

As in the case of spin systems \cite{rigol06,rigol06a}, the fundamental 
basis of our numerical linked cluster expansion, 
for some extensive property $P$ of an infinite lattice ${\cal L}$, 
is the relation \cite{domb,book}
\begin{equation}
P({\cal L})/N=\sum_c L(c)\times W_P(c),
\label{dirsum}
\end{equation}
where the left hand side is the value of the property $P$ per lattice site 
in the thermodynamic limit. On the right hand side $L(c)$ is the so-called 
lattice constant that is the number of embeddings of the cluster $c$, 
per lattice site, in the lattice ${\cal L}$. $W_P(c)$ is the weight of
the cluster $c$ for the property $P$. The latter is defined recursively 
by the principle of inclusion and exclusion \cite{domb}
\begin{equation}
W_P(c)=P(c)-\sum_{s\subset c}W_P(s),
\label{weights}
\end{equation}
where $P(c)$ is the property $P$ calculated for the finite cluster $c$. 
The sum on $s$ runs over all subclusters of $c$. The basic idea of NLC 
is to calculate $P$ and $W_P$ at any temperature by means of an exact 
diagonalization of each cluster $c$. 

For itinerant models, it is desirable to control the density in the 
thermodynamic limit. Within NLC this is achieved by working in the 
grand canonical ensemble, i.e., by introducing a chemical potential. 
Hence, in this case
\begin{eqnarray}
\label{pc1} 
P(c)&=& \dfrac{1}{Z_c}\mathrm{Tr}\left\lbrace 
P\ \exp\left[ -\left( {\cal H}-\mu\sum_{i} n_i\right) /T\right]\right\rbrace_c.\ \
\end{eqnarray}
In Eq.\ (\ref{pc1}), $\mu$ denotes the chemical potential,  
$T$ is the temperature of the system (we have set the Boltzmann constant 
$k_B$ to be unity), and $Z_c$ is the partition function in each cluster $c$
\begin{equation}
\label{Z} 
Z_c=\mathrm{Tr} \left\lbrace 
\exp\left[-\left( {\cal H}-\mu\sum_{i} n_i\right) /T\right]\right\rbrace_c .
\end{equation}
Notice that in Eq.\ (\ref{pc1}) and Eq.\ (\ref{Z}), $\mathrm{Tr}$ denotes
the grand canonical trace, i.e., we fully diagonalize the $t$-$J$ Hamiltonian
for all possible fillings in each cluster. This, together with the fact that 
the Hilbert space is larger (three states per site) for the $t$-$J$ model, 
when compared with $1/2$-spin models (with only two states per site) studied 
in Refs.\ \cite{rigol06,rigol06a}, reduces the size of the largest clusters we 
can consider here. All results presented in this paper correspond to the site 
expansion of the square lattice (see Ref.\ \cite{rigol06a} for details) 
with all clusters up to 10 sites. At zero doping, i.e., the Heisenberg
model, we present NLC results considering all clusters with up to 13 sites 
\cite{rigol06,rigol06a}. In addition, throughout this work we examine
two values of the parameter $J/t$: $J/t=0.5$ as in the HTE studies of 
Ref.\ \cite{singh92}, and $J/t=0.3$ as in the FTLM studies of 
Refs.\ \cite{jaklic00,jaklic94}, to both of which we compare 
some of our results. 

\begin{figure}[!htb]
\begin{center}
  \includegraphics[scale=1.3,angle=0]{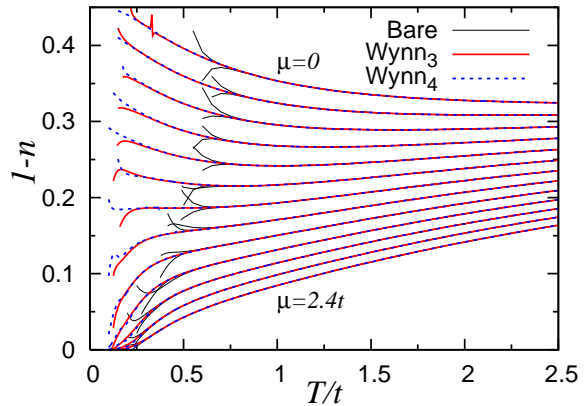}
\end{center}
\vspace{-0.5cm}
\caption{\label{HoleDensvsTemp}
(Color online) Hole density as a function of temperature for different 
chemical potentials (from top to bottom $\mu=0,0.2,\ldots,2.4$), 
and $J/t=0.5$. For each value of the chemical potential we have plotted 
results of the bare NLC sums up to 9 and 10 sites (thin continuous lines) 
and results of Wynn extrapolations after three and four cycles of improvement 
(see Ref.\ \cite{rigol06a} for details).}
\end{figure}

For itinerant models, the first quantity one needs to evaluate 
within NLC is the density. This is because one is, in general, interested
in the behavior of thermodynamic observables as a function of the
temperature at a fixed density or as a function of the density 
(or hole doping) at a fixed temperature. In  Fig.\ \ref{HoleDensvsTemp}
we plot the hole density as a function of the temperature for different 
chemical potentials. The results of the bare NLC sums (\ref{dirsum}) 
exhibit a clear feature. They converge to lower temperatures as 
the density approaches 1.

In order to access lower temperatures we use the 
sequence extrapolation techniques detailed in Ref.\ \cite{rigol06a}.
Results of Wynn extrapolations for the density after three and four 
cycles of improvement are also depicted in Fig.\ \ref{HoleDensvsTemp}. 
For the lowest chemical potential ($\mu=0$) one can see that series 
extrapolations allow one to extend the region of convergence of the 
density from $T/t>0.75$ to $T/t<0.2$. 

\begin{figure}[!htb]
\begin{center}
  \includegraphics[scale=1.3,angle=0]{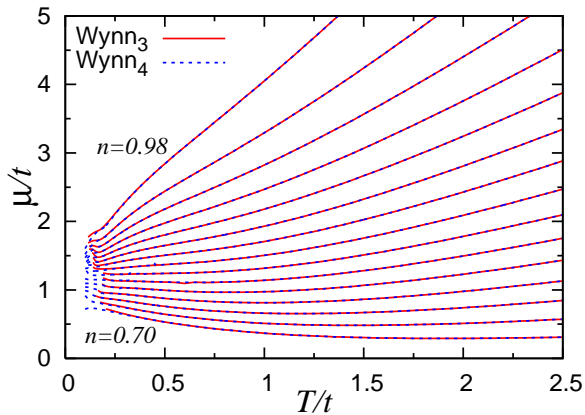}
\end{center}
\vspace{-0.5cm}
\caption{\label{ChemPotvsTemp}
(Color online) Chemical potential as a function of temperature for 
different densities (from bottom to top $n=0.7,0.72,\ldots,0.98$), 
and $J/t=0.5$. For each value of the density we have plotted 
results of Wynn extrapolations after three and four cycles of improvement.}
\end{figure}

Once we have performed the extrapolations for the density, we numerically 
invert the dependence of any observable from $(\mu,T)$ to $(n,T)$. 
As a first example we show in Fig.\ \ref{ChemPotvsTemp} the dependence 
of the chemical potential on the temperature when the density is held 
fixed. As can be seen, it is possible to follow the chemical potential 
curves at constant density to rather low temperatures $T/t\sim 0.1$.

\begin{figure}[!htb]
\begin{center}
  \includegraphics[scale=1.3,angle=0]{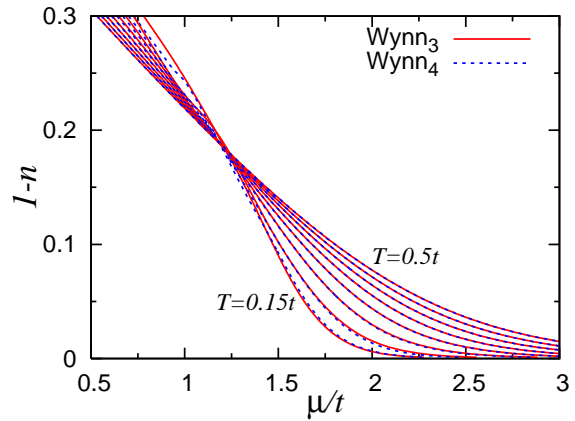}
\end{center}
\vspace{-0.5cm}
\caption{\label{HoleDensvsChemPot}
(Color online) Hole density vs chemical potential for 
different temperatures (below the crossing point 
from bottom to top $T/t=0.15,0.20,\ldots,0.5$), 
and $J/t=0.5$. For each value of the temperature we have 
plotted results of Wynn extrapolations after three and four cycles 
of improvement.}
\end{figure}

An analysis of the results depicted in Fig.\ \ref{ChemPotvsTemp} for the 
lowest temperatures suggests that the dependence of $\mu$ on the temperature 
is roughly linear
\begin{equation}
\mu(T)=\mu(T=0)+ A\ T.
\end{equation}  
This is in agreement with the results obtained by JP 
\cite{jaklic00} using FTLM, and departs from the $T^2$ dependence expected 
of a Fermi liquid. In addition, there is a change in the slope $A$ 
as the density is shifted. For $J/t=0.5$ we find that the slope $A$ changes 
sign when $n'\sim 0.81$. This can be better seen in 
Fig.\ \ref{HoleDensvsChemPot},
where we plot the hole density vs the chemical potential at fixed
temperatures. There, one can see that for $1-n'\sim 0.19$ the different 
curves cross each other. For $1-n' < 0.19$ our NLC results are 
well converged at all temperatures $T/t>0.15$. Small departures between 
Wynn extrapolations after three and four cycles of improvement are only apparent
for the lowest temperatures and highest hole concentrations.

We have also performed a similar study for $J/t=0.3$. In this case we 
have obtained, in close agreement with JP \cite{jaklic00}, that the 
change of slope in $A$ occurs for $n'\sim 0.86$.

\section{Entropy \label{entropy}}

Once one has numerically inverted the dependence ($\mu,T$) to 
($n,T$) it is possible to study the behavior of observables of interest 
as a function of the temperature at a fixed density. In this section we
consider the entropy per lattice site
\begin{equation}
S=\frac{1}{N}\left( \ln Z + 
\frac{ \left\langle{\cal H}-\mu\sum_{i} n_i \right\rangle}{T} 
\right),
\end{equation} 
where the term proportional to the chemical potential 
on the right hand side, not present in our 
calculations for spin systems, is needed when dealing with the 
grand canonical ensemble. $N$ is the number of lattice sites.

\begin{figure}[!htb]
\begin{center}
  \includegraphics[scale=1.3,angle=0]{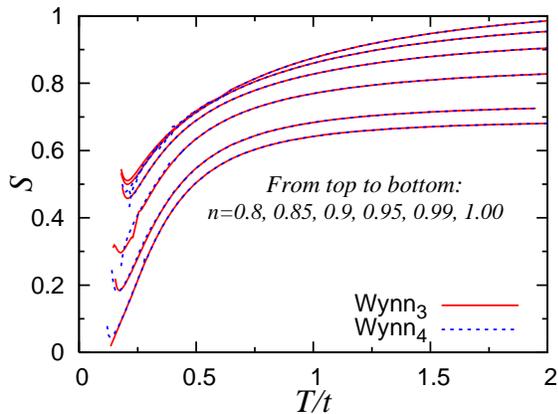}
\end{center}
\vspace{-0.5cm}
\caption{\label{EntropyvsTemp}
(Color online) Entropy as a function of temperature for 
different densities and $J/t=0.5$. For each value of the density 
we have plotted results of Wynn extrapolations after three and four 
cycles of improvement.}
\end{figure}

In Fig.\ \ref{EntropyvsTemp}, we plot the entropy vs temperature for
different values of the density after three and four cycles of improvements
using Wynn's algorithm. As with the chemical potential, the NLC 
convergence for $S$ improves as the density approaches 1. For all the 
densities plotted in Fig.\ \ref{EntropyvsTemp} we get good convergence
for $T/t\geq 0.25$.

\begin{figure}[!htb]
\begin{center}
  \includegraphics[scale=1.3,angle=0]{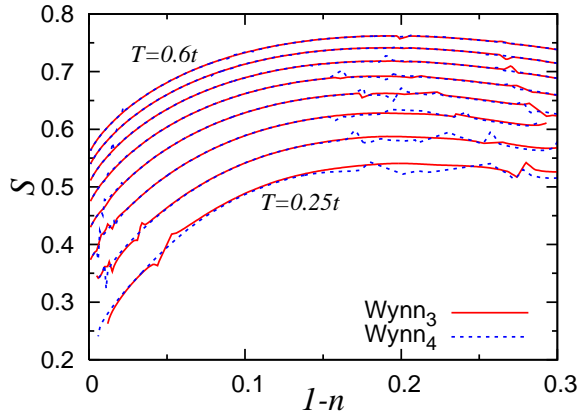}
\end{center}
\vspace{-0.5cm}
\caption{\label{EntropyvsHoleDens}
(Color online) Entropy as a function of the density for 
different temperatures 
(from bottom to top $T/t=0.25,0.20,\ldots,0.6$) 
and $J/t=0.5$. For each value of the temperature 
we have plotted results of Wynn extrapolations after three and four 
cycles of improvement.}
\end{figure}

Now we discuss the dependence of the entropy on density 
when the temperature is held fixed. This is shown in 
Fig.\ \ref{EntropyvsHoleDens}. The entropy exhibits
a very broad maximum that slowly shifts towards lower hole 
densities as the temperature is increased. These results are
in qualitative agreement with experimental measurements that
have found the entropy to be maximum around $1-n=0.22$ 
at a temperature $T/t=0.07$ (lower than the ones we have 
calculated here) \cite{loram96}.

\begin{figure}[!htb]
\begin{center}
  \includegraphics[scale=0.43,angle=0]{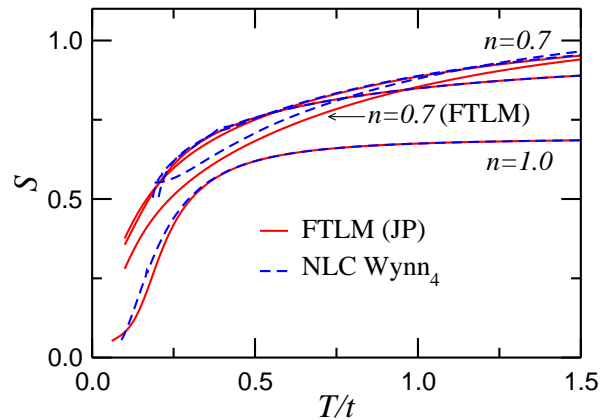}
\end{center}
\vspace{-0.5cm}
\caption{\label{Entropy}
(Color online) Entropy as a function of temperature for 
different densities (from top to bottom at $T/t=1.5$ NLC results 
correspond to $n=0.7,0.8,0.9,$\ and\ 1.0) and $J/t=0.3$. 
Our Wynn extrapolations after four cycles of improvement are 
compared with FTLM results by JP \cite{jaklic00}.}
\end{figure}

To conclude this section on the entropy we compare in 
Fig.\ \ref{Entropy} our NLC results with those obtained by JP 
\cite{jaklic00} using FTLM in clusters with 20 lattice sites and 
$J/t=0.3$. At zero doping (Heisenberg model), the agreement between 
NLC (with clusters up to 13 sites \cite{rigol06}) and FTLM 
is remarkable down to $T/t\sim 0.25$. Below that temperature, finite 
size effects start to be apparent as FTLM results marginally depart 
from the ones obtained using NLC (see also discussion in 
Ref.\ \cite{rigol06} on the entropy of the Heisenberg model on 
the square lattice). 

Within FTLM, finite size effects for the entropy, at a given temperature, 
increase when holes are added to the antiferromagnet. This can be 
understood within the analysis presented in Ref.\ \cite{jaklic00}. 
There it was shown that the temperature at which finite size effects 
start to become relevant increases with doping $1-n\gtrsim 0.15$, 
for the system sizes considered in that work. In particular, for 
$n=0.7$ (the worst case for FTLM) our NLC results for $S$ are 
slightly different from the ones in Ref.\ \cite{jaklic00} for all 
temperatures in Fig.\ \ref{Entropy}.

\section{Uniform susceptibility \label{suscept}}

We study in this section another thermodynamic quantity of much
experimental interest, the uniform susceptibility
\begin{equation}
\chi = \frac{\left\langle \left( S^z_{\textrm{tot}}\right)^2\right\rangle}{NT}.
\end{equation}

\begin{figure}[!htb]
\begin{center}
  \includegraphics[scale=1.3,angle=0]{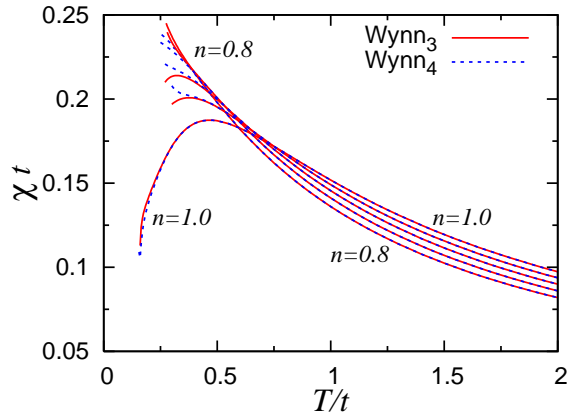}
\end{center}
\vspace{-0.5cm}
\caption{\label{SusceptibilityvsTemp}
(Color online) Uniform susceptibility as a function of the temperature 
for different densities (from bottom to top for the highest 
temperatures $n=0.8,0.85,\ldots,1.0$) and $J/t=0.5$. 
For each value of the density we have plotted results of Wynn 
extrapolations after three and four cycles of improvement.}
\end{figure}

In Fig.\ \ref{SusceptibilityvsTemp} we show results for the uniform
susceptibility as a function of temperature for different densities.
For the Heisenberg model, the results of our Wynn 
extrapolations for $\chi$ are well converged down to $T/J\sim0.3$. 
They allow one to resolve the peak in $\chi$ that occurs 
around $T/J\sim 1$, and signals the onset of short range 
antiferromagnetic order. With doping the peak shifts to lower 
temperatures, i.e., the doping slows down the growth of 
antiferromagnetic correlations, and eventually disappears. 
Only for $n=0.9$ and 0.95 do our NLC results exhibit signals 
of a peak in $\chi$.

\begin{figure}[!htb]
\begin{center}
  \includegraphics[scale=1.3,angle=0]{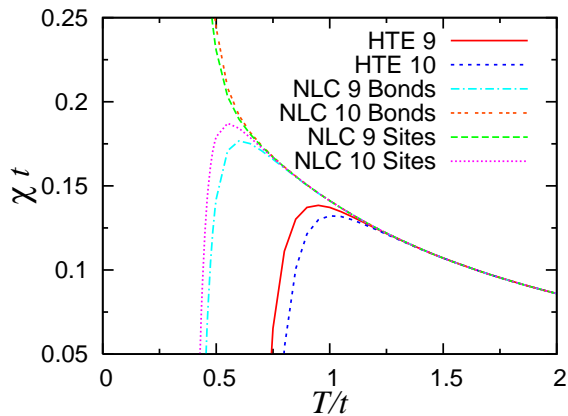}
\end{center}
\vspace{-0.5cm}
\caption{\label{SusceptibilityvsTemp_NLCvsHTE_B}
(Color online) Uniform susceptibility results for the bare NLC sums 
of the bond and site expansions compared with those obtained with HTE up 
to order 10. The density in the system is $n=0.85$.}
\end{figure}

In what follows we compare NLC results for the uniform susceptibility
with the ones obtained using HTE \cite{singh92}. We first contrast, 
in Fig.\ \ref{SusceptibilityvsTemp_NLCvsHTE_B}, the results of the 
NLC bare sums (\ref{dirsum}) for the bond and site expansions 
with those obtained for the HTE bare sums up to order 10.  
Figure \ref{SusceptibilityvsTemp_NLCvsHTE_B} shows that the direct NLC 
sums converge down to $T/t\sim 0.75$ (with the site expansion being
slightly better than the bond expansion), while the HTE results are 
only well converged to $T/t\sim 1.25$. The existence of this region 
of temperatures where NLC converges while HTE does not was our main 
motivation for developing NLC, and makes NLC a more controlled 
technique at intermediate temperatures. 

\begin{figure}[!htb]
\begin{center}
  \includegraphics[scale=1.3,angle=0]{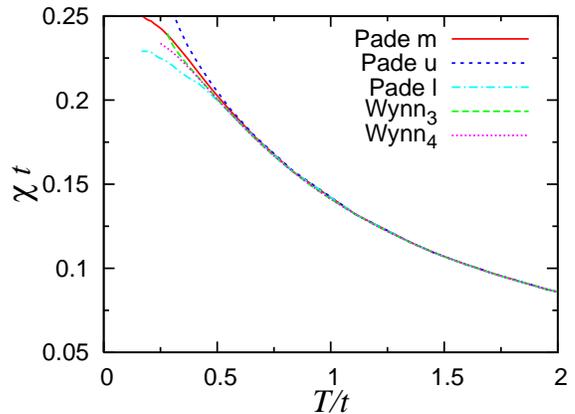}
\end{center}
\vspace{-0.5cm}
\caption{\label{SusceptibilityvsTemp_NLCvsHTE_E}
(Color online) Pade approximants for the uniform susceptibility
\cite{singh92} are compared with Wynn extrapolations of the NLC site
expansion. In the legend, Pade m, Pade u, and Pade l, indicate the 
mean estimated value, upper, and lower limits, respectively, obtained 
by different Pade approximants. The density in the system is $n=0.85$.}
\end{figure}

As discussed before, an important feature of NLC and 
HTE is that both approaches allow for systematic extrapolations
that accelerate the convergence of NLC and enable going beyond 
the radius of convergence of HTE. In 
Fig.\ \ref{SusceptibilityvsTemp_NLCvsHTE_E} we compare Pade 
extrapolations for $\chi$ with Wynn extrapolations for the NLC 
site expansion. The agreement between these two approaches is 
remarkable down to $T/t\sim 0.5$, which show that indeed 
extrapolations can work quite well for both techniques,
if the temperature is not too low. For this 
quantity, however, different Wynn extrapolations converge to lower 
temperatures than the Pade approximants. We find this to be remarkable 
for itinerant models because the analytic structure of HTE allows 
for an analytic inversion of the grand-canonical ($\mu,T$) dependence 
into the more experimentally relevant one ($n,T$). Within NLC we have to 
perform a numerical extrapolation to access lower temperatures and 
then do a numerical inversion ($\mu,T$)$\rightarrow$($n,T$).
Still, the fact that NLC contains exact information from the finite
clusters at all temperatures, while in HTE only a power series
expansion in inverse temperature is kept, means that the former,
in a real sense, requires less extrapolation.

\section{Specific heat \label{cv}}

In this section we analyze the NLC results for the specific heat 
and compare them with those obtained within FTLM 
\cite{jaklic00}. The specific heat is defined as 
\begin{equation}
C_v= T \left( \frac{\partial S}{\partial T}\right)_{N_e}.
\end{equation}
However, we avoid the numerical differentiation of the entropy
with respect to the temperature by evaluating
\begin{eqnarray}
C_v= \frac{\langle {\cal H}^2 \rangle-\langle {\cal H} \rangle^2}{NT^2}-\frac{1}{NT}\left(\frac{\partial N_e}{\partial \mu} \right)_T
\left\lbrace \left(\frac{\partial{\langle\cal H\rangle}} 
{\partial N_e} \right)_T\right\rbrace ^2,\ \ 
\end{eqnarray}
which substantially reduces the numerical errors as the NLC 
sums for the number of particles ($N_e=\langle\sum_{i}n_i\rangle$) 
and the energy ($\langle\cal H\rangle$) converge (and are better 
behaved) to lower temperatures than the ones of the entropy. 
Still, we have to perform numerical derivatives in addition to 
calculating the fluctuations of the energy 
($\langle {\cal H}^2 \rangle-\langle {\cal H}\rangle^2$). Hence, 
of the thermodynamic observables analyzed 
so far $C_v$ is the most difficult for NLC evaluation.

\begin{figure}[!htb]
\begin{center}
  \includegraphics[scale=1.3,angle=0]{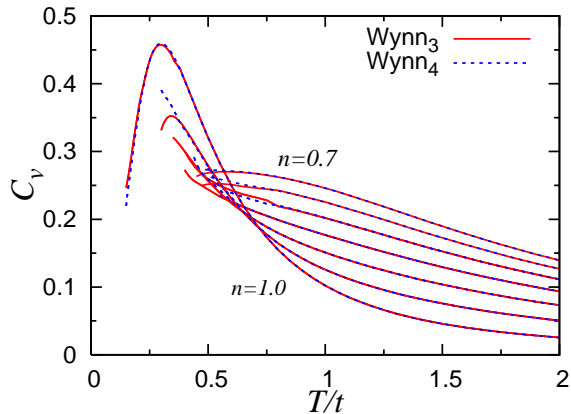}
\end{center}
\vspace{-0.5cm}
\caption{\label{SpecificHeatvsTemp}
(Color online) Specific heat as a function of the temperature 
for different densities (from top to bottom 
$n=0.7,0.75,\ldots,1.0$) and $J/t=0.5$. 
For each value of the density we have plotted results of Wynn 
extrapolations after three and four cycles of improvement.}
\end{figure}

\begin{figure}[!htb]
\begin{center}
  \includegraphics[scale=1.3,angle=0]{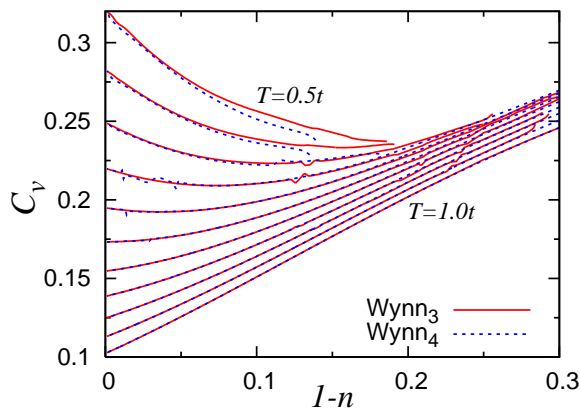}
\end{center}
\vspace{-0.5cm}
\caption{\label{SpecificHeatvsHoleDens}
(Color online) Specific heat as a function of the hole density
for different temperatures (from top to bottom 
$n=0.5,0.55,\ldots,1.0$) and $J/t=0.5$. For each value of the 
temperature we have plotted results of Wynn extrapolations 
after three and four cycles of improvement.}
\end{figure}

In Fig.\ \ref{SpecificHeatvsTemp} we plot the specific heat as a function
of the temperature for different densities. For $C_v$ we only obtain 
well converged results below $T/t=0.5$ for the Heisenberg model and 
the $t$-$J$ model at low hole concentration. Figure \ref{SpecificHeatvsTemp} 
shows that the maximum in the specific heat, attributed to the thermal 
activation of the spin degrees of freedom, becomes strongly suppressed
with doping. This can be better seen in Fig.\ \ref{SpecificHeatvsHoleDens}, 
where we have plotted the specific heat as a function of the hole density 
for fixed values of the temperature. For the lowest temperatures in that 
figure, we only can follow the $C_v$ curves for doping below 0.15. They 
clearly show that $C_v$ decreases as the doping is increased, 
and exhibits a minimum that moves towards higher doping concentrations 
as the temperature is lowered. 

\begin{figure}[!htb]
\begin{center}
  \includegraphics[scale=0.43,angle=0]{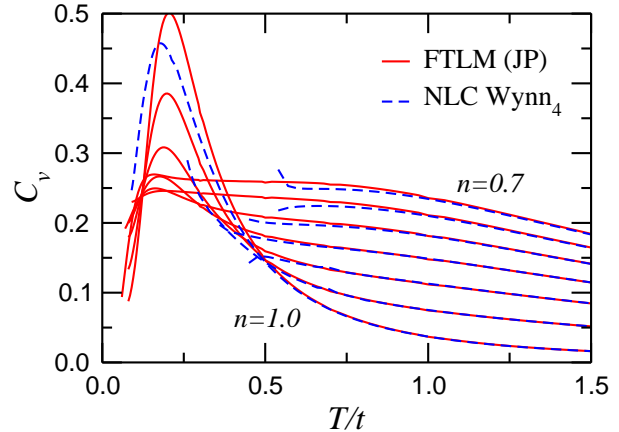}
\end{center}
\vspace{-0.5cm}
\caption{\label{Cv}
(Color online) Specific heat as a function of the temperature 
for different densities (from top to bottom 
$n=0.7,0.75,\ldots,1.0$) and $J/t=0.3$. 
Our Wynn extrapolations after four cycles of improvement are 
compared with FTLM results by JP \cite{jaklic00}.}
\end{figure}

In Ref.\ \cite{rigol06a} we have argued that the specific heat of 
the Heisenberg model suffers from large finite size effects at 
relatively high temperatures. (We presented results for the full 
diagonalization of a $4\times4$ lattice.) In Fig.\ \ref{Cv} we compare 
our NLC results for the $t$-$J$ model with the ones obtained by 
JP for clusters with 20 lattice sites and $J/t=0.3$. At high 
temperatures $T/t>0.75$ our NLC results are in very good agreement 
with the ones of JP. NLC results for $C_v$, however, do not converge 
below $T=0.5$ when the doping in the system is large. As the density 
approaches 1, and the convergence of NLC moves to lower temperatures, 
our linked-cluster results start to depart from the ones obtained with
FTLM. In general, we find our calculation of the specific heat in the 
thermodynamic limit to be below the FTLM results for finite clusters. 
This deviation becomes particularly large for the Heisenberg model. 
The FTLM peak for $C_v$ when $N=20$ occurs at larger temperatures 
than the one obtained within NLC. Interestingly, increasing the system size
up to $N=26$ does not help since the peak remains at the same position 
while becoming higher \cite{jaklic00}, instead of moving to lower
temperatures and becoming smaller as has been found to be the case 
in the thermodynamic limit \cite{rigol06a,bernu01}. This suggests that
in general, specific heat is a more difficult quantity to calculate
numerically.  

\section{Conclusions \label{conclusions}}

We have presented an application of the recently 
introduced NLC approach to itinerant models. In particular,
we have studied thermodynamic properties of the $t$-$J$ model 
on the square lattice.

From the possible NLC expansions discussed in Ref.\ \cite{rigol06a},
we have found that the one best suited to the $t$-$J$ model on the
square lattice is the site based expansion. For this expansion we
have shown that the NLC bare sums (for the observables considered 
here) converge to lower temperatures than the bare HTE sums. In 
addition, Wynn extrapolations for the site based expansion were
found to provide better results at lower temperatures than the 
ones obtained with Pade extrapolations for HTE. This in spite
of the fact that the inversion of the dependence ($\mu,T$) to 
($n,T$) can be done analytically within HTE, while within NLC
it is done numerically.

Through a comparison with results obtained by Jakli\v c and Prelov\v sek 
using the finite-temperature Lanczos method \cite{jaklic00,jaklic94}, 
we have also shown that NLC allows one to access regions at low temperatures 
where finite size effects are relevant to exact diagonalization studies. 
One particularly striking example presented here is the peak in the specific 
heat of the Heisenberg model. In the calculations done in 
Refs.\ \cite{jaklic00,jaklic94} it was found that with increasing the system 
size (within the sizes that can be addressed by exact diagonalization) 
the peak does not move while it does increase in size. NLC results, which 
confirmed previous results by Bernu and Misguich \cite{bernu01}, 
exhibit a smaller peak shifted towards lower temperatures.

Moving away from Heisenberg models, it remains to be seen as to 
which method is in general accurate to lower temperatures. Future 
application of NLC to superconducting and other exotic susceptibilities 
should prove informative. The study of the former using HTE remains
controversial \cite{pryadko,ogata,putikka}. It may also prove useful 
to combine NLC with Lanczos and FTLM methods to extend it to still 
lower temperatures.

\begin{acknowledgments}

This work was supported by the US National Science Foundation, 
Grant Nos.\ DMR-0240918, DMR-0312261, and PHY-0301052. We are 
grateful to P. Prelov\v sek for providing us with the data from 
Ref.\ \cite{jaklic00}, and for useful comments on FTLM.

\end{acknowledgments}

\end{document}